\documentclass[12pt]{article}

\usepackage[dvips]{color}                 
\usepackage{graphicx}                     
\usepackage{amssymb}
\usepackage{amsmath}
\usepackage{xspace}                       
\newcommand{\units}[1]{\ensuremath{\,\mathrm{#1}}}

\begin{document}

\newcommand{\mytitle}[1]{\begin{center} \LARGE{\textbf{#1}} \end{center}}
\newcommand{\myauthor}[1]{\textbf{#1}}
\newcommand{\myaddress}[1]{\textit{#1}}
\newcommand{\mypreprint}[1]{\begin{flushright} #1 \end{flushright}}

\begin{titlepage}
	\mypreprint{\textbf {TUM/T39-06-10} \par}
	\vspace*{1.0cm}
	\mytitle{Chiral extrapolation of $g_A$ with explicit $\Delta\,(1232)$ degrees of freedom}
	\vspace*{1cm}
	\begin{center}
		\myauthor{M.~Procura},
		\myauthor{B.~U.~Musch},
		\myauthor{T.~R.~Hemmert}
		and
		\myauthor{W.~Weise} \par
		\vspace*{0.5cm}
		\myaddress{Physik-Department, Theoretische Physik,  \\
			Technische Universit{\"a}t M{\"u}nchen, D-85747 Garching, Germany\\
			}
		\vspace*{0.2cm}
		\end{center}
	\vspace*{2.5cm}

\begin{abstract}
An updated and extended analysis of the quark mass dependence of the nucleon's axial vector coupling constant $g_A$ is presented in comparison with state-of-the-art lattice QCD results. Special emphasis is placed on the role of the $\Delta(1232)$ isobar. It is pointed out that standard chiral perturbation theory of the pion-nucleon system at order $p^4$ fails to provide an interpolation between the lattice data and the physical point. In constrast, a version of chiral effective field theory with explicit inclusion of the $\Delta(1232)$ proves to be successful. Detailed error analysis and convergence tests are performed. Integrating out the $\Delta(1232)$ as an explicit degree of freedom introduces uncontrolled errors for pion masses $m_\pi \gtrsim 300\,{\rm MeV}$.	
\end{abstract}
	
	\end{titlepage}
	
\setcounter{page}{2}
\newpage

\sloppy
 
\section{Introduction}

The axial-vector coupling constant $g_A$ of the nucleon represents a benchmark test of our ability to extract hadron properties from the QCD Lagrangian. Its empirical value is accurately determined from neutron $\beta$-decay \cite{PDG04}: $g_A=1.267 \pm 0.003$. On the other side, both lattice QCD calculations \cite{RBCK,negele,newQCDSF} and chiral effective field theories \cite{BFHM,HPW,ulf} are making progress in describing the quark mass dependence of this nucleon property. 

The present paper updates and extends our previous study in Ref.\,\cite{HPW} about chiral extrapolations for $g_A$, in the continuum and infinite volume limit. 
We refer to \cite{Tim,newQCDSF,bs} for chiral perturbation theory analyses of the effects due to the finite spatial extent of the lattice simulation volume \cite{GLfv}.
In Ref.\,\cite{HPW} we compared two different two-flavor, non-relativistic chiral effective field theories at leading-one-loop level: Heavy Baryon Chiral Perturbation Theory (HB$\chi$PT), with pions and nucleons as active degrees of freedom, and the so-called Small Scale Expansion (SSE), which includes the $\Delta\,(1232)$ explicitly. Treating the $\Delta\,(1232)$ as an explicit degree of freedom turned out to be crucial in order to obtain a consistent extrapolation of available lattice data down to the region of small quark masses. 

The important role played by the $\Delta\,(1232)$ in the physics behind $g_A$ does not come as a surprise. It has in fact been known for decades that the $\Delta$-dominance of $P$-wave pion-nucleon scattering, or equivalently, the strong spin-isospin polarizability of the nucleon, has its pronounced impact on matrix elements of the axial current in the nucleon ground state. The Adler-Weisberger sum rule is a prominent example illustrating this connection. Extended versions of chiral effective field theories, such as SSE, have been designed to incorporate these well established features.

A surprising outcome of lattice QCD results for $g_A$ is their weak dependence on the quark mass. This suggests a subtle balance between contributions from the different degrees of freedom involved. In fact, substantial cancellations between $\pi N$ and $\pi \Delta$ loops in $g_A$ have been reported first by Jenkins and Manohar \cite{JM2}.
In Section \ref{ganumpiN} we demonstrate that no interpolation between physical point and state-of-the-art lattice results is possible, with parameters consistent with hadron phenomenology, if we use standard chiral perturbation theory with only pions and nucleons up to next-to-leading-one-loop order. This observation holds both for the non-relativistic and infrared regularized, manifestly covariant scheme \cite{BL}. Section \ref{AWsec} recalls the Adler-Weisberger sum rule and outlines the importance of the $\Delta\,(1232)$ for $g_A$ in a simple schematic model. 
In Section \ref{gassenum} we present a detailed statistical analysis showing that the leading-one-loop SSE expression with explicit $\Delta\,(1232)$ worked out in Ref.\,\cite{HPW} does lead to successful interpolations with recent lattice results even for relatively large values of $m_\pi$. 
Remarkably, the outcome of our fits  turns out to be consistent with available phenomenological information about the low-energy couplings involved.
In Section \ref{gasseexp} we explore the mapping between expressions worked out with and without explicit $\Delta\,(1232)$ by examining the separate contributions of different powers of $m_\pi$ in the expanded SSE result. 
Conclusions are drawn in Section \ref{conclusions}.

Part of this study has been anticipated in several conference proceedings, see Refs.\,\cite{baryons,cyprus,panic}.

\section{$g_A$ without explicit $\Delta\,(1232)$ up to ${\cal O}(p^4)$} \label{ganumpiN}

\begin{figure}[t]
  \begin{center}
    \includegraphics*[width=0.7\textwidth]{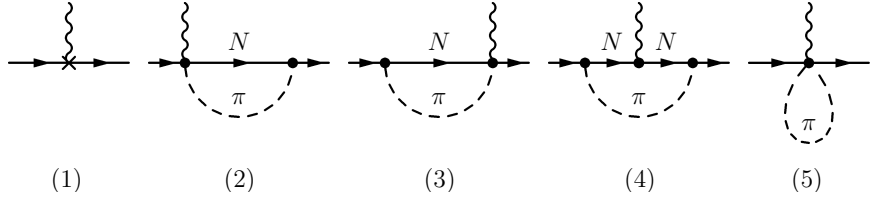}
     \caption{Diagrams contributing to the quark mass dependence of $g_A$ up to order $p^3$ in Baryon Chiral Perturbation Theory. The wiggly line denotes an external isovector axial-vector field, interacting with a nucleon (solid line). The first graph to the left encodes the relevant counterterms.}\label{diaggap3}
  \end{center}
\end{figure}

\begin{figure}[t]
  \begin{center}
    \includegraphics*[width=0.16\textwidth,clip=true,trim=21.5 0 110 0]{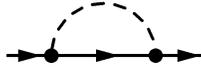}
     \caption{Graph contributing to nucleon field renormalization at order $p^3$. }\label{gaZcube}
  \end{center}
\end{figure}

The axial-vector coupling constant $g_A$ of the nucleon is defined as the limit of the nucleon axial form factor at vanishing momentum transfer $Q^2$. 
It is extracted from the forward (on-shell) nucleon matrix element of the isovector axial-vector quark current $\bar{q} \gamma_\mu \gamma_5 (\tau^i/2)\,q$ where $q$ denotes the $(u, d)$ isospin doublet quark field and $\tau^i$ are Pauli matrices. 

In this section we point out the incapability of standard chiral perturbation theory, with only pion and nucleon as explicit degrees of freedom, to deal with the quark mass dependence of $g_A$.
The formalism has been worked out up to next-to-leading one-loop order, ${\cal O}(p^4)$, both in HB$\chi$PT \cite{KaMo} and in the manifestly covariant framework employing infrared regularization \cite{julia}. For the sake of completeness we summarize in Appendix \ref{appgaBChPT} technical details of the calculations with infrared regularization, from which the HB$\chi$PT expressions can also be extracted.

The diagrams relevant at leading-one-loop level, ${\cal O}(p^3)$, are drawn in Figs.\,\ref{diaggap3} and \ref{gaZcube}.
In those loop graphs all vertices are extracted from the leading $\pi N$ Lagrangian 
\begin{equation}
{\cal{L}}_{\pi N}^{(1)}=\frac{g_A^0}{2}\, \bar{\Psi} \left(-\frac{1}{f_\pi}\, \vec{\tau} \cdot \partial_\mu\vec{\pi} +2 a_\mu\right) \gamma^\mu \gamma_5 \,\Psi  + \dots \label{Luno}
\end{equation}
where $\Psi$ is the nucleon field; $g_A^0$ and $f_\pi$ denote, respectively, the nucleon axial coupling and the pion decay constant in the $SU(2)$ chiral limit, {\it i.e.} for vanishing $u$- and $d$-quark masses. Here $a_\mu=a_\mu^i \tau^i/2$ represents an external isovector axial field. 

\begin{figure}[t]
  \begin{center}
    \includegraphics*[width=0.35\textwidth]{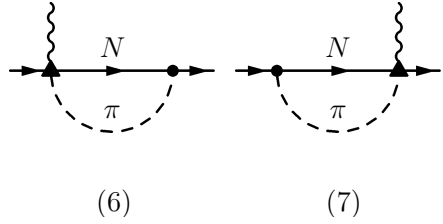}
     \caption{Diagrams contributing to $g_A$ at the next-to-leading one-loop level. The triangle denotes a vertex appearing in ${\cal{L}}_{\pi N}^{(2)}$. The wiggly line represents an external axial vector field.}\label{garelpfourth}
  \end{center}
\end{figure}

The next-to-leading-one-loop calculation involves vertices from the second order pion-nucleon Lagrangian:
\begin{equation}
{\cal L}_{\pi N}^{(2)} = 4\, c_1 m_\pi^2\, \bar{\Psi} \Psi+ \frac{c_3}{2}\, {\rm Tr}(u_\mu u^\mu) \bar{\Psi} \Psi - \frac{c_4}{4} \,\bar{\Psi}\, \gamma^\mu \gamma^\nu [u_\mu, u_\nu] \Psi + \dots~ \label{Ldue}
\end{equation}
where
\begin{equation}
u_\mu = -\frac{1}{f_\pi}\, \vec{\tau} \cdot \partial_\mu \vec{\pi} +2 a_\mu+ \dots
\end{equation}
The trace is taken over the isospin indices. In writing down the first term of Eq.\,(\ref{Ldue}) we have already used the connection between pion- and $u$-, $d$-quark masses given by the Gell-Mann-Oakes-Renner relation, neglecting isospin violating effects.

The ${\cal O}(p^4)$ result for $g_A(m_\pi)$ is obtained by evaluating the diagrams in Fig.\,\ref{garelpfourth} and the contributions from wave function (Fig.\,\ref{gaZfourth}) and mass renormalization, see Appendix \ref{appgaBChPT}. The former ones contain the pion-nucleon-axial vertex from ${\cal L}_{\pi N}^{(2)}$ with the two low-energy constants $c_3$ and $c_4$, which are known to primarily encode the influence of the $\Delta\,(1232)$ resonance on low-energy pion-nucleon dynamics (see for example the discussion in Ref.\,\cite{PMWHW}). 

\begin{figure}[t]
  \begin{center}
    \includegraphics*[width=0.15\textwidth,clip=true,trim=34 0 110 0]{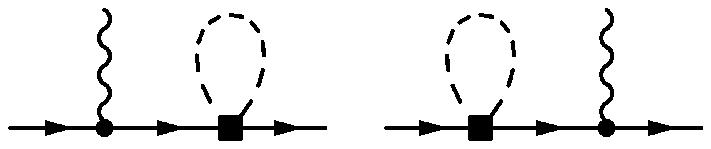}
     \caption{Graph contributing to the nucleon $Z$-factor at order $p^4$. The square denotes a vertex from ${\cal{L}}_{\pi N}^{(2)}$ involving $c_1$, $c_2$ and $c_3$. } \label{gaZfourth}
  \end{center}
\end{figure}

The expansion of the next-to-leading one-loop expression, Eq.\,(\ref{gap4}), around the chiral limit gives
\begin{align}
g_A=\;g_A^0&+\left[4\,C(\lambda)-\frac{{g_A^0}^3}{16 \pi^2 f_\pi^2}-\frac{g_A^0+2{g_A^0}^3}{8 \pi^2 f_\pi^2}\ln{\frac{m_\pi}{\lambda}}\right] m_\pi^2+ \left(\frac{g_A^0+{g_A^0}^3}{8\pi f_\pi^2 M_0}+\frac{2c_4-c_3}{6 \pi f_\pi^2}\right)m_\pi^3 \nonumber\\& +\left[\frac{4g_A^0(1+2{g_A^0}^2)+
g_A^0M_0(3 c_2+16c_4)}{64 \pi^2 f_\pi^2 M_0^2} +32\,F(\lambda) \right.\nonumber\\& \left.+\,\frac{g_A^0[2+3{g_A^0}^2-M_0(3c_2 +4c_3 -4c_4)]}{16 \pi^2 f_\pi^2 M_0^2}\ln{\frac{m_\pi}{\lambda}}\right]m_\pi^4+ \dots \label{gap4chir} 
\end{align}
Here $C(\lambda)$ and $F(\lambda)$ are effective couplings representing unresolved short-distance dynamics and compensating logarithmic scale ($\lambda$-) dependence.
The sum of the terms up to and including $m_\pi^3$ coincides with the ${\cal{O}}(p^4)$ expression in dimensionally regularized HB$\chi$PT \cite{KaMo}. 

\subsection{Numerical analysis}

We analyse leading and next-to-leading one-loop expressions in comparison with an updated set of lattice QCD results for the quark mass dependence of $g_A$, provided by the RBCK \cite{RBCK}, LHPC \cite{negele} and QCDSF \cite{newQCDSF}  collaborations. 

\begin{figure}[t]
  \begin{center}
    \includegraphics[width=0.7\textwidth]{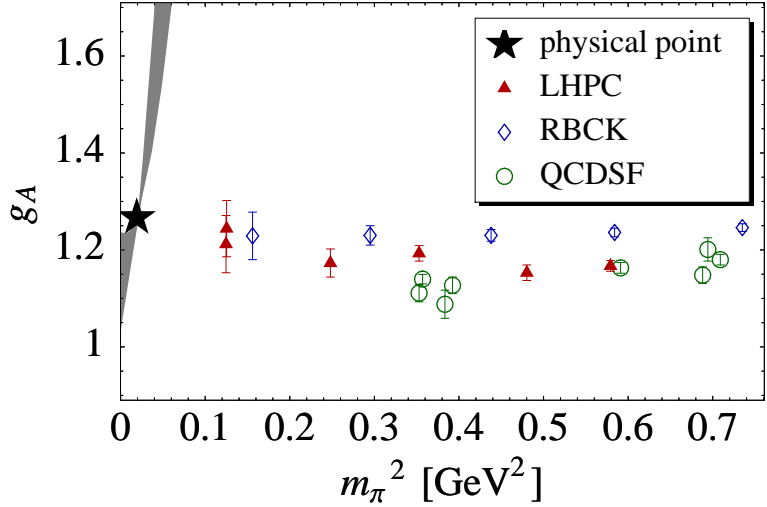}
    \caption{Data base of lattice QCD results for the pion mass dependence of $g_A$ provided by the RBCK \cite{RBCK}, QCDSF \cite{newQCDSF} and LHPC \cite{negele} collaborations. Also shown is the ${\cal O}(p^4)$ result for $g_A(m_\pi)$ in Heavy Baryon $\chi$PT, with the physical point included as a constraint. The band reflects the uncertainty on the input values of the low-energy constants involved.}
    \label{bandnodelta}
  \end{center}
\end{figure}

Fig.\,\ref{bandnodelta} summarizes these sets of data, although they refer to simulations with different actions, number of flavors, lattice volumes, spacings and procedures to translate the results into physical units. The reasons for the evident discrepancies between different lattice groups are not yet fully understood, see Ref.\,\cite{newQCDSF}. Among the data in Ref.\,\cite{RBCK} we have selected those produced on the larger lattice, with spatial size  $L= 2.4\,{\rm{fm}}$. The lowest-$m_\pi$ LHPC point corresponds to $L=3.5\,{\rm fm}$, see \cite{negele}. 
Both the LHPC and QCDSF collaborations performed full-QCD simulations whereas the RBCK data are quenched.
Preliminary, unquenched RBCK simulations \cite{RBCKbis} still have too large error bars to make definite statements in the region of interest.

As already mentioned, we translate the quark mass $m_q$-dependence of $g_A$ into a pion mass dependence according to the Gell-Mann-Oakes-Renner relation, the leading order linear relation between $m_\pi^2$ and $m_q$ in Chiral Perturbation Theory. Recent accurate lattice QCD results clearly display this behavior for a wide range of quark masses \cite{Lusch,Deb,Gock,Aoki}, even for $m_\pi > 0.6\,{\rm GeV}$, for reasons not yet understood in detail.

In a first step the quark mass dependence of $g_A$ in chiral perturbation theory is analysed {\it without fitting to the available lattice data}. We have produced Monte Carlo bands both from Eq.\,(\ref{gap3}) and Eq.\,(\ref{gap4}) and from the Heavy Baryon formulae at order $p^3$ and $p^4$. In doing this we eliminate $g_A^0$ by imposing the physical constraint $g_A(m_\pi^{\rm phys})=1.267$. The remaining parameters are randomly chosen within phenomenologically acceptable ranges. 
The dimension-two low-energy constants $c_1$, $c_2$, $c_3$ and $c_4$ are constrained, within non-relativistic chiral effective field theory for pions and nucleons, by several low-energy $\pi N$ and $N N$ scattering studies. Combining results of Refs.\,\cite{BKMpiN}, \cite{BM} and \cite{EM} and simply superimposing the quoted error bars, we obtain the following ranges, in units of ${\rm GeV}^{-1}$:
\begin{equation}
c_1 \approx -1.0 \dots -0.7~,\;\;\;c_2 \approx 3.1 \dots 3.5~,\;\;\;c_3 \approx -5.6 \dots -3.4~,\;\;\;c_4 \approx 3.4 \dots 3.7~.
\end{equation}
The value of $c_1$ basically drives the pion-nucleon sigma term. For a detailed discussion on the value of $c_3$, see Ref.\,\cite{PMWHW}.
The analysis in Refs.\,\cite{nadia1,nadiathesis} of $\pi N\rightarrow \pi\pi N$ scattering at leading-one-loop order in HB$\chi$PT limits the combination of couplings $C^r(\lambda)$ in Eq.\,(\ref{ccc}). According to \cite{nadiathesis}, $C^r(\lambda=m_\pi^{\rm phys})=-1.4 \pm 1.2\,{\rm GeV}^{-2}$.
For the higher-order couplings in the infrared regularized expressions --- called $F$ and $G$ in Appendix \ref{appgaBChPT} --- we rely on ``naive'' dimensional arguments \cite{GM}, at a regularization scale $\lambda=1\,{\rm GeV}$ : $F^r(1\units{GeV})=(-1 \ldots 1)\units{GeV^{-4}}$, $G^r(1\units{GeV})=(-1 \ldots 1)\units{GeV^{-6}}$. For the nucleon mass in the chiral limit we scan the range $M_0=(0.88 \ldots 0.89)\units{GeV}$ according to the outcome of the analysis in Refs. \cite{PMWHW,PHW}.

None of the Monte Carlo bands for $g_A$, which include uncertainties of the low-energy parameters, comes anywhere close to the lattice data in the $m_\pi$ region of interest. This is demonstrated in Fig.\,\ref{bandnodelta} for the case of the ${\cal O}(p^4)$ Heavy Baryon expression. What numerically drives the trend of the quark mass dependence of $g_A$ at order $p^4$ is the combination of $c_3$ and $c_4$ in the term proportional to $m_\pi^3$ in Eq.\,(\ref{gap4chir}) \footnote{The large contribution associated with the $m_\pi^3$ term was already pointed out in Ref.\,\cite{KaMo}.}. Recoil corrections to the non-relativistic results which one gets in the infrared regularization approach, do not improve the situation.

We can conclude at this point that chiral perturbation theory of the pion-nucleon system, at order $p^4$ and without explicitly propagating $\Delta\,(1232)$, fails in the attempt to provide an interpolation of $g_A$ between the physical point and lattice QCD results. The price one would have to pay for enforcing adjustment of an interpolating curve to the lattice data is that the combination $2 c_4 - c_3$ of $\pi N$ low-energy constants must be tuned to a value totally inconsistent with the empirical ones. For chiral perturbation theory to be able to make contact with present lattice data, compensation of the strong $m_\pi^3$ trend must come from higher powers of $m_\pi$. In a recent paper \cite{ulf} a relatively flat quark mass dependence is shown to be possibly achieved at two-loop level in the Heavy Baryon framework. In this approach the compensation of the $m_\pi^3$ term does not arise from the ``double log'' piece characteristic of the two-loop calculation but from a fine-tuning of unknown effective couplings at fifth order. The theoretical uncertainties are reported to be acceptably small only for $m_\pi \lesssim 300\,{\rm MeV}$. 

A more efficient and physically motivated way to successfully interpolate between lattice results and the physical point is to include explicit $\Delta\,(1232)$ degrees of freedom \cite{HPW}.\footnote{An extrapolation for $g_A(m_\pi)$ using a chiral quark model with $\Delta(1232)$ contributions has been produced in Ref. \cite{DET}.}
This leads to a whole string of higher powers in $m_\pi$ already at leading-one-loop level and the outcome of the fits agrees favourably with available information from low-energy hadron phenomenology, see Section \ref{gassenum}. 
Intermediate $\Delta\,(1232)$ contributions are well-known to play a crucial role in axial current matrix elements because the axial-vector field induces strong isovector $N({1/2}^+) \to \Delta({3/2}^+)$ transitions \cite{TW}. The near degeneracy of the $\Delta\,(1232)$ with the nucleon suggests treating both $N$ and $\Delta$ as explicit degrees of freedom. Before focusing on the detailed formalism in Section \ref{deltasection}, it is useful to recall, in the next section, some well-known basic physics which outlines the special relevance of the $\Delta\,(1232)$ in the present context of $g_A$.

\section{Adler-Weisberger sum rule and a schematic model} \label{AWsec}

The Adler-Weisberger (AW) sum rule \cite{AW} combines information from low-energy QCD, current algebra and dispersion relations to connect low-energy parameters of the $\pi N$ system ($f_\pi$ and $g_A$) with an integral over the difference of the $\pi^{+} p$ and $\pi^{-} p$ total cross-sections.
\begin{figure}[t]
 \begin{center}
 \includegraphics*[width=.48\linewidth]{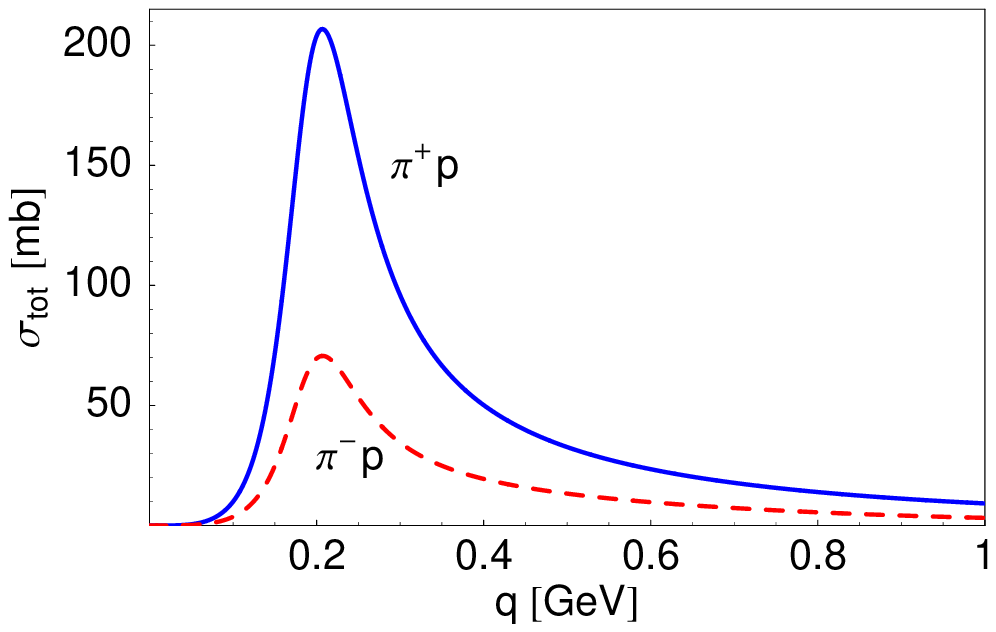}
 \includegraphics*[width=.48\linewidth]{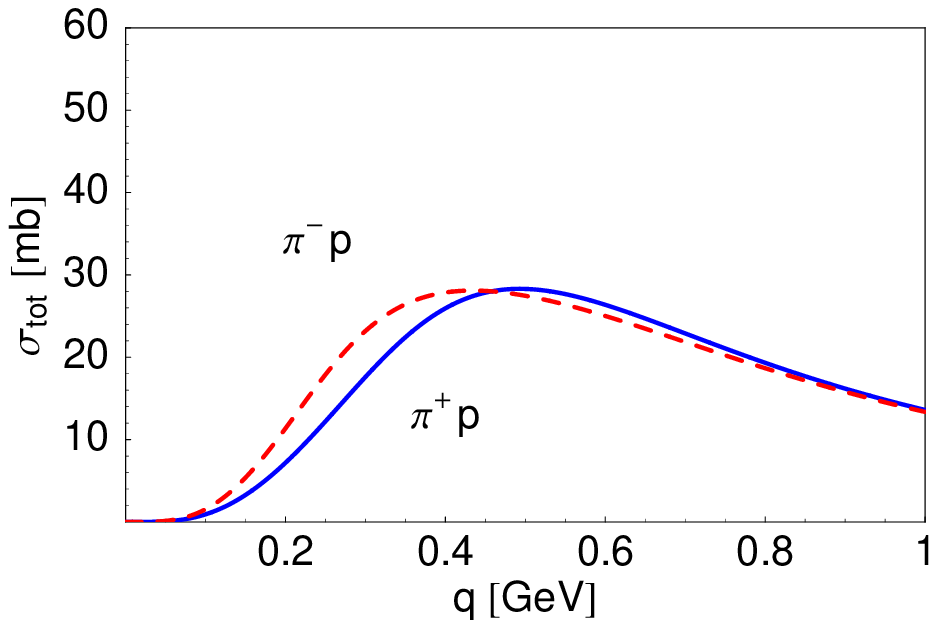}
  \caption{$P$-wave contributions from the nucleon and $\Delta\,(1232)$ pole graphs to the total $\pi^+ p$ and $\pi^- p$ cross sections, plotted against the pion momentum in the laboratory frame. In the left panel the direct and crossed $\Delta\,(1232)$-pole graphs are included, in the right panel not.} \label{AWfig}
 \end{center}
\end{figure}
Using the Goldberger-Treiman relation and the Weinberg-Tomozawa low-energy theorem \cite{WT} one obtains{\footnote{ See, for example, the derivation in \cite{TW}.}}
\begin{equation}
g_A^2=1+\frac{2 f_\pi^2}{\pi}\int_{0}^{\infty}\frac{d q}{\omega}\,[\sigma_{\pi^+ p}(\omega)-\sigma_{\pi^- p}(\omega)]+{\cal O}\left(\frac{m_\pi^2}{M_N^2}\right)~, \label{aw} 
\end{equation}
where the integral is taken over the pion momentum, $q=\left|{\vec{q}\,}\right|=\sqrt{\omega^2-m_\pi^2}$, and $\omega$ denotes the pion energy in the nucleon rest frame. The deviation of $g_A$ from 1 is tied to pion-nucleon dynamics and spontaneous (and explicit) chiral symmetry breaking. The left- and right-hand sides of the sum rule turn out to agree at the percent level using $f_\pi=92.4\,{\rm{MeV}}$ and an accurate parameterization of the measured $\pi^{\pm} p$ cross-sections \cite{ELT}.

In order to demonstrate the prominent role of the $\Delta\,(1232)$ in the AW sum rule for $g_A$ it is instructive to perform a simple schematic model calculation as follows. Consider the $P$-wave $\pi N$ forward scattering amplitudes $f_\alpha(\omega)$ in spin-isospin channels $\alpha=(2I, 2J)$. Their contributions to the total $\pi N$ cross section is
\begin{equation}
\sigma_{\alpha}(\omega)=\frac{4 \pi}{\left|{\vec{q}\,}\right|}\,{\rm Im} f_\alpha(\omega)
\end{equation}
according to the optical theorem. Next, introduce the $K$-matrix $K_\alpha$ in each channel by
\begin{equation}
f_\alpha(\omega)=\frac{K_\alpha(\omega)}{1-i \left|{\vec{q}}\,\right|\,K_\alpha(\omega)}~. \label{vbn}
\end{equation}
$K_\alpha$ has poles on the real $\omega$-axis located at the physical masses of the corresponding $N$ and $\Delta$ intermediate states.

Starting from the $\pi N N$ vertex Eq.\,(\ref{Luno}) and the leading-order $\pi N \Delta$ transition Lagrangian
\begin{equation}
{\cal L}_{\pi N \Delta}^{(1)}= -\frac{c_A}{f_\pi}\, \bar{\Psi}^i_\mu \,\partial^\mu \pi^i\, \Psi + {\rm h.~c.}~,
\end{equation}
where $\Psi^i_\mu$ is the Rarita-Schwinger field of the $\Delta$, the $K$-matrix pieces involving direct and crossed $N$ and $\Delta$ pole terms are, in the non-relativistic limit \cite{EW}:
\begin{align}
K_{3 3}&=\frac{\vec{q}\,^2}{12 \pi f_\pi^2}\left[\frac{g_A^2}{\omega}+\frac{c_A^2}{\Delta-\omega}+\frac{1}{9} \frac{c_A^2}{\Delta+\omega}\right]~, \nonumber \\
K_{1 1}&=\frac{\vec{q}\,^2}{3 \pi f_\pi^2}\left[-\frac{g_A^2}{2 \omega}+\frac{4}{9} \frac{c_A^2}{\Delta+\omega}\right]~, \nonumber \\
K_{1 3}&= K_{3 1} = \frac{1}{4} K_{1 1}~, \label{Kii}
\end{align}
where $\Delta= M_\Delta- M_N$ is the delta-nucleon mass splitting deduced from the position of the resonance pole in the complex center-of-mass energy plane.

We use $f_\pi = 92.4\units{MeV}$ on the right-hand sides of these last equations, together with $\Delta=271.1\,{\rm MeV}$, the delta pole position determined empirically from magnetic dipole and electric quadrupole transition amplitudes \cite{HDT}, and $c_A=1.5$ from the $\Delta\rightarrow\pi N$ decay width (see for example Ref.\,\cite{PMWHW}). Then the $P$-wave $\pi N$ scattering volumes deduced from Eq.\,(\ref{Kii}) at $\omega=m_\pi$ agree very well with experiment, see also Ref.\,\cite{EW}. Eq.\,(\ref{vbn}) and the optical theorem lead to the $P$-wave contributions to the total $\pi^+ p$ and $\pi^- p$ cross-sections: 
\begin{equation}
\sigma_{\pi^+ p}=\frac{4 \pi}{\left|{\vec{q}\,}\right|}\,{\rm Im} \left[2 f_{3 3}+ f_{3 1}\right]~,\;\;\;\;\;\;\;\sigma_{\pi^- p}=\frac{4 \pi}{3 \left|{\vec{q}\,}\right|}\,{\rm Im} \left[2 f_{3 3}+ f_{3 1}+ 4 f_{1 3}+2 f_{1 1}\right]~.
\end{equation}
They are drawn in the left panel of Fig.\,\ref{AWfig}. 
The curves are in good agreement with the empirical ones from near threshold up to $q \approx 0.5\,{\rm GeV}$ \cite{ELT, EW}, while the region between $0.5$ and $2\,{\rm GeV}$ receives contributions from higher resonances. Using the cross-sections in Eq.\,(\ref{aw}), we obtain $g_A=1.22$. The integral on the right-hand side of the Adler-Weisberger sum rule is indeed dominated by the contribution of the $\Delta\,(1232)$ to the $\pi N$ scattering amplitude. If the calculation includes only the nucleon Born terms from Eqs.\,(\ref{Kii}), setting $c_A=0$, $\sigma_{\pi ^{\pm} p}$ change as shown in the right panel of Fig.\,\ref{AWfig}. ``Switching off" the $\pi N \Delta$ coupling, the contribution from the dispersion integral in the Adler-Weisberger sum rule changes sign and $g_A$ is reduced to $0.99$.

This non-relativistic, tree-level calculation illustrates the role of the $\Delta(1232)$ at the physical pion mass. In our previous paper \cite{HPW} we have also shown that treating the $\Delta(1232)$ as an explicit degree of freedom leads to successful chiral extrapolation for pion masses well above $m_\pi^{\rm phys}$. The lattice results used there are now outdated by more recent and improved sets. In the next section we check whether our conclusions in Ref.\,\cite{HPW} remain unaltered for the most recent input data and perform a detailed statistical analysis, using the same methods as described in Ref.\,\cite{PMWHW}.

\begin{figure}[t]
  \begin{center}
    \includegraphics*[width=0.7\textwidth]{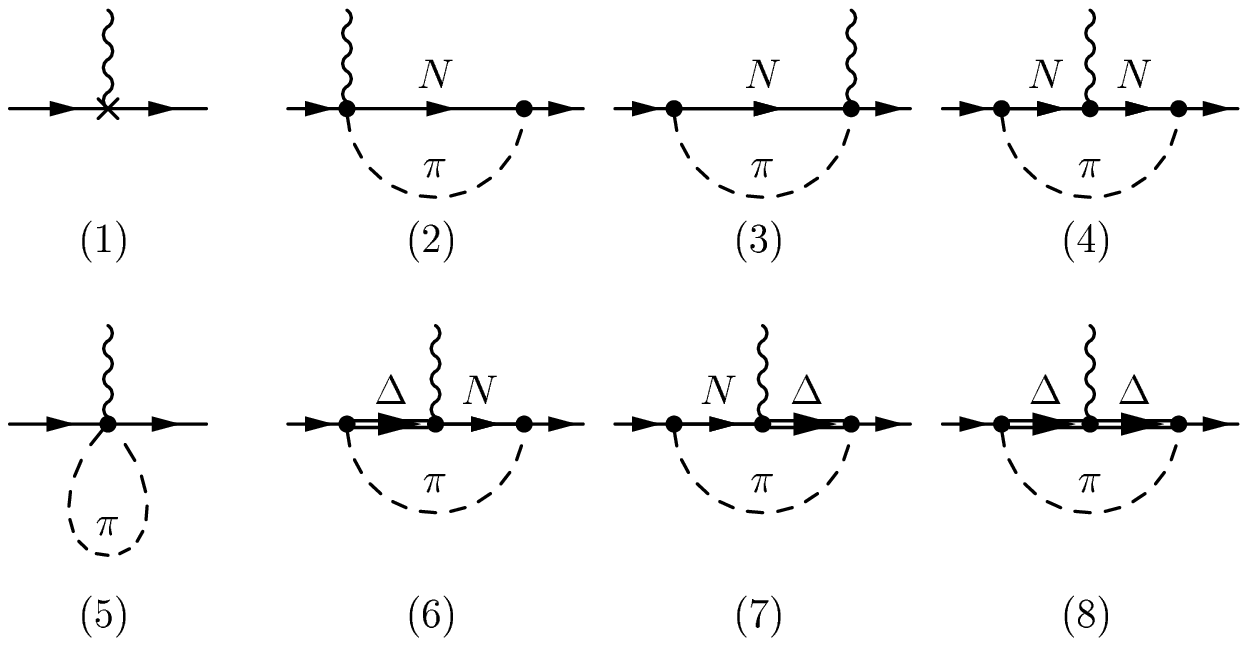}
    \caption{$g_A$ in the Small Scale Expansion at leading-one-loop order. The wiggly line denotes an external isovector axial-vector field. All vertices shown here appear in the leading $\pi N$ and $\pi N \Delta$ Lagrangians.}
    \label{gadiags}
  \end{center}
\end{figure}

\begin{figure}[t]
  \begin{center}
    \includegraphics*[width=0.16\textwidth,clip=true,trim=22 0 110 0]{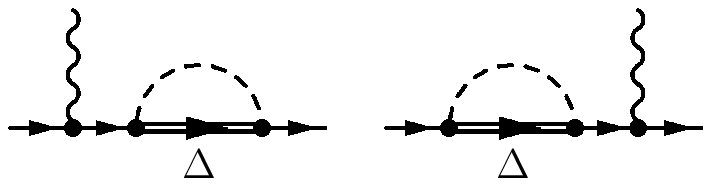}
    \caption{Graph contributing to wave function renormalization at order $\epsilon^3$.}
    \label{gaZeps}
  \end{center}
\end{figure}

\section{$g_A$ with explicit $\Delta\,(1232)$} \label{deltasection}
In the framework of the so-called Small Scale Expansion (SSE) \cite{HHK}, the delta-nucleon mass splitting $\Delta$ is treated as a small parameter and included in the power counting of ``small scales'' generically denoted by $\epsilon$.  The leading one-loop, order $\epsilon^3$, contribution to $g_A$ is represented by the diagrams in Figs.\ref{gadiags} and \ref{gaZeps}. We recall here the result in the continuum and infinite volume case with dimensional regularization \cite{HPW}:
\begin{align}
g_A^{\rm SSE}(m_\pi^2)=g^0_A&-\frac{{g^0_A}^3\,m_\pi^2}{16\pi^2f_\pi^2}
                     +4\left[C^{\rm SSE}(\lambda)
                     +\frac{c_A^2}{4\pi^2 f_\pi^2}\left(\frac{155}{972}\, g_1
                     -\frac{17}{36}\, g^0_A\right)+\gamma^{\rm SSE}
                     \ln{\frac{m_\pi}{\lambda}}\right]m_\pi^2\nonumber\\
                  & +\frac{4c_A^2\, 
                     g^0_A}{27 \pi f_\pi^2\, \Delta}\,m_\pi^3
                     +\frac{8}{27\pi^2 f_\pi^2}\;c_A^2\, g^0_A\, m_\pi^2\,
                     \sqrt{1-\frac{m_\pi^2}{\Delta^2}}\,\ln{R}\nonumber\\
                  & +\frac{c_A^2\Delta^2}{81\pi^2 f_\pi^2}\left(25g_1-
                     57g_A^0\right)\left(\ln\frac{2\Delta}{m_\pi}
                     -\sqrt{1-\frac{m_\pi^2}{\Delta^2}}\,\ln R\right)
                     +{\cal O}(\epsilon^4)\;,\label{gasse}
\end{align}
with
\begin{align}
\gamma^{\rm SSE}&=\frac{-1}{16\pi^2 f_\pi^2}\left[g_A^0\left({1\over 2}+{g_A^0}^2\right) + \frac{2}{9}\,c_A^2\left(g_A^0-\frac{25}{9}g_1\right)\right]\;,\nonumber\\
R&=\frac{\Delta}{m_\pi }+\sqrt{\frac{\Delta^2}{m_\pi^2}-1}\,.\label{R}
\end{align}
$C^{\rm SSE}(\lambda)$ is a combination of renormalized third-order couplings, cf. Eq.\,(\ref{ccc}), and $g_1$ is the axial delta-delta coupling; $f_\pi$ and $\Delta$ are the $SU(2)$ chiral limit values of the pion decay constant and the delta nucleon mass splitting, respectively. See Ref.\,\cite{HPW} for further details. The analytic continuation of the previous expression for $m_\pi > \Delta$ is achieved via the replacement
\begin{equation}
\sqrt{\Delta^2 - m_\pi^2}\,\ln{\left(\sqrt{\frac{\Delta^2}{m_\pi^2}-1}+\frac{\Delta}{m_\pi}\right)}\, \to\, - \sqrt{m_\pi^2 - \Delta^2}\, \arccos{\left(\frac{\Delta}{m_\pi}\right)}~.
\end{equation}
Note that in Eq.\,(\ref{gasse}) decoupling of the delta has been implemented up to working order \cite{newQCDSF,HPW}. As a result, $g_A^0$ is the same coupling as in HB$\chi$PT calculations without explicit delta. This is different in Ref.\,\cite{bs}. 

\subsection{Fit results} \label{gassenum}
We have performed fits based on Eq.\,(\ref{gasse}) and its analytic continuation for $m_\pi > \Delta$ both to the two-flavor, quenched RBCK data and the full-QCD, 2 (lighter) +1 (heavier)-flavor LHPC results. From the RBCK data we have selected the three points with the lightest pion masses. The values of $f_\pi$, $c_A$ and $\Delta$ have been fixed as input. Without any loss of generality we have set the regularization scale $\lambda$ equal to $1\,{\rm GeV}$. The effective coupling $C^{\rm SSE}(\lambda=1\,{\rm GeV})$ has been eliminated by imposing that the fit curves pass through the physical point.

\begin{figure}[t]
  \begin{center}
    \includegraphics*[width=0.47\textwidth]{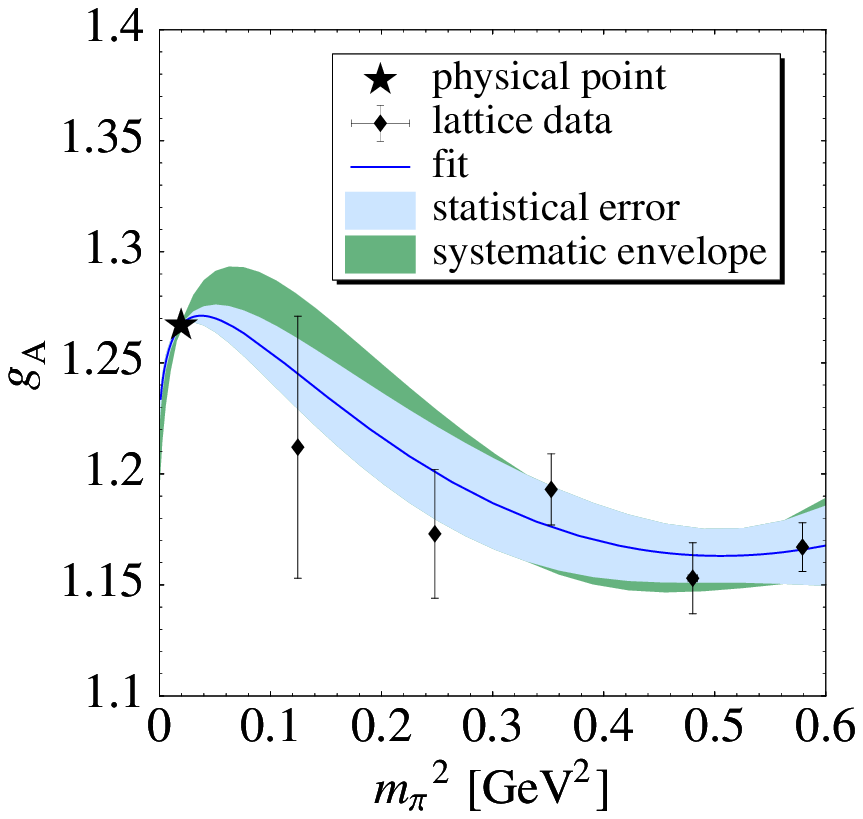}
    \includegraphics*[width=0.47\textwidth]{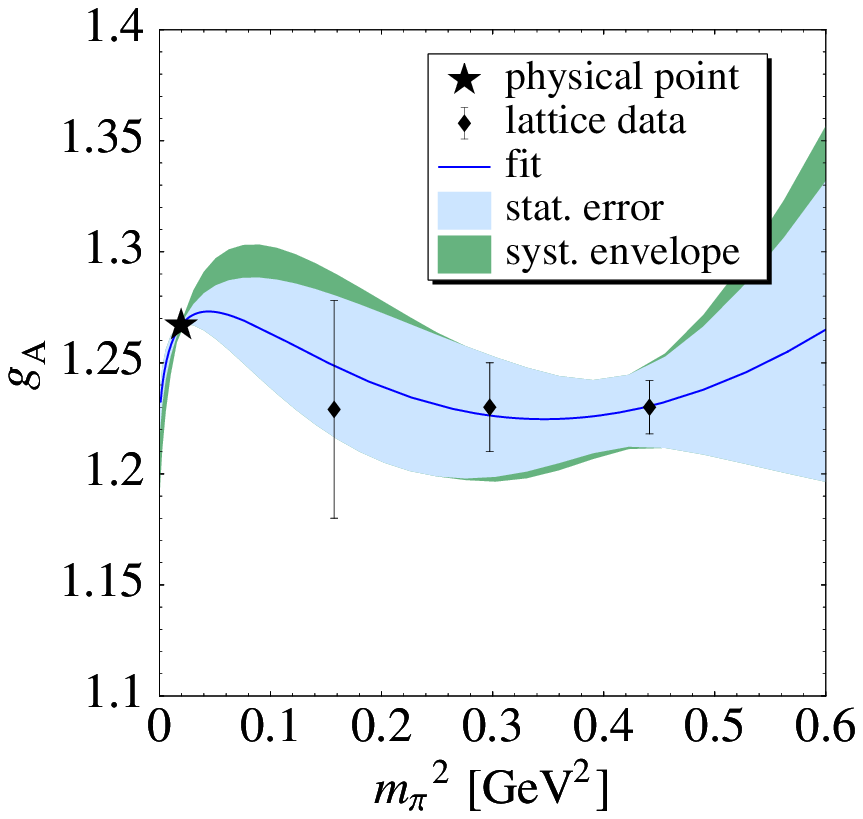}
    \caption{${\cal O}(\epsilon^3)$ SSE best-fit curves, $68\%$-statistical error bands and systematic envelopes for the LHPC lattice results \cite{negele} (left) and RBCK data \cite{RBCK} (right). The physical point is included as a constraint. See Table \ref{tableLHPC}.}
    \label{bandLHPC}
  \end{center}
\end{figure}


\begin{table*}[tb]
	\caption{Fit results for the ${\cal O}(\epsilon^3)$ SSE interpolation between the physical value of $g_A$ and the LHPC points \cite{negele}.
	}
	\label{tableLHPC}
	\renewcommand{\arraystretch}{1.2}
\begin{tabular}{ll||cc|cc} 
	 & & \multicolumn{2}{l|}{(a) statistical error} & \multicolumn{2}{l}{(b) systematic envelope}  \\
	\hline
	$g_A^0$ &                                             & 1.224 $\pm$ 0.004 & fitted & 1.191 $\dots$ 1.228 & fitted \\   
	$g_1$ &                                               & 2.80 $\pm$ 0.12 & fitted & 2.6 $\dots$ 6.0 & fitted \\
	$C^\text{SSE}(1 \units{GeV})$ & $(\mathrm{GeV^{-2}})$ & $-1.65$ $\pm$ 0.17 & elim. & $-4.3$ $\dots$ $-1.4$ & elim. \\     
	$f_\pi$ & $(\mathrm{MeV})$                            & 92.4 & fixed  & 86.2 $\dots$ 92.4 & scanned \\    
	$c_A$ &                                               & 1.5 & fixed & 1.12 $\dots$ 1.5 & scanned \\         
	$\Delta$ & $(\mathrm{MeV})$                           & 271.1 & fixed & 271 $\dots$ 293 & scanned \\            
	\hline
	$\chi^2/\text{d.o.f.}$ &                              & 0.93 & & 0.92 $\dots$ 1.27 & \\         
	\hline
	$c_A^2 g_1$ &                                         & 6.29 $\pm$ 0.27 & & 5.9 $\dots$ 7.6 & \\  
	$C^\text{HB}(m_\pi^\text{phys})$ & $(\mathrm{GeV^{-2}})$ & $-0.45$ $\pm$ 0.05 & & $-0.55$ $\dots$ $0.41$ &  \\ 
	\end{tabular}
\end{table*}

Fig.\,\ref{bandLHPC} and Table \ref{tableLHPC} summarize our results for the LHPC data. In drawing the 1-$\sigma$ error band (which takes into account correlations between the parameters), we set $f_\pi$ equal to its physical value and $c_A=1.5$. Furthermore, $\Delta=271.1\,{\rm MeV}$ from the real part of the complex delta pole in the total center-of-mass energy plane \cite{HDT}. 

The ``systematic band'' \cite{PMWHW} in Fig.\,\ref{bandLHPC} quantifies the sensitivity to variations of the input parameters. It is given by the envelope of the 1-$\sigma$ bands scanning also the additional input values $f_\pi=86.2\,{\rm MeV}$ \cite{CD}, $c_A=1.125$ \cite{HPW} and $\Delta= 293\,{\rm MeV}$ (from the $90^0$ $\pi N$ phase-shift in the spin-3/2 isospin-3/2 channel \cite{PDG04}). We point out that the constant $g_1$ appears only in combinations multiplied by $c_A^2$. This is evident from diagram (8) of Fig.\,\ref{gadiags}. In the fit using $c_A=1.125$, the apparent strong deviation of $g_1$ from the $SU(4)$ quark model prediction $g_1=9 g_A /5 \approx 2.2$ is therefore of little relevance. The range accessible to $c_A^2 \, g_1$ is indeed much smaller than that for $g_1$ itself, see Table \ref{tableLHPC}. \par

Our estimate of $C^{\rm SSE}(1\, {\rm GeV})$ is consistent with (limited) information from $\pi N \to \pi \pi N$ scattering. One can indeed link $C^{\rm HB}(\lambda)$ in the Heavy Baryon $\pi N$ effective field theory and $C^{\rm{SSE}}(\lambda)$ in the framework with explicit delta. Comparing the terms proportional to $m_\pi^2$ in the chiral expansion of Eq.\,(\ref{gasse}) and Eq.\,(\ref{gap3chir}), 
one gets:
\begin{equation}
C^{\rm HB}(\lambda)=C^{\rm SSE}(\lambda)+\frac{c_A^2}{72\,\pi^2\,f_\pi^2}\left[5\left(\frac{23}{27}g_1-\frac{7}{3}g_A^0\right)+\left(\frac{25}{9}g_1-g_A^0\right)\ln{\frac{2 \Delta}{\lambda}}\right] \label{Cpar}~.
\end{equation}
According to our fits we obtain $C^{\rm HB}(\lambda=m_\pi^{\rm phys})=(-0.45 \pm 0.05)\,{\rm GeV}^{-2}$, consistent with the broad range of values extracted from Ref.\,\cite{nadiathesis}.
In agreement with \cite{HPW}, we stress that the ``chiral log" in the leading non-analytic quark-mass term is only visible for pion masses {\em well below} the physical point.

We have also performed fits to the three LHPC points below $600\,{\rm MeV}$ in pion mass only. The outcome is compatible, within large error bars, with our previous results for $m_\pi \lesssim 760\,{\rm MeV}$ in Table \ref{tableLHPC}.

Table \ref{tableRBCK} summarizes our study of the RBCK data. The error bars of the output parameters absorb the effects of one heavier flavor, together with the above mentioned systematic discrepancies due to different fermion actions, (un)quenching, and translation of lattice results into physical units.

Given the present data situation, extrapolations without the physical point as input need to include finite volume dependence in order to get an estimate of $g_A$ at the physical pion mass with reasonable statistical accuracy. Those studies have been performed by the LHPC \cite{negele} and the QCDSF collaborations \cite{newQCDSF}, the latter one using our SSE scheme of Ref. \cite{HPW} extended to finite volume. The couplings determined in Ref. \cite{newQCDSF} agree with our fit results of Table\,\ref{tableLHPC}. Remarkably, both collaborations obtain extrapolated values of $g_A$ at the physical pion mass that are consistent with phenomenology, although the present level of accuracy needs to be improved. 

In the future, reliable chiral extrapolations will certainly be based on combined fits of several observables, including finite volume corrections. Such a program will become feasible once simulations at smaller pion masses will be available.


\begin{table*}[tb]
	\caption{Fit results for the ${\cal O}(\epsilon^3)$ SSE interpolation between the physical value of $g_A$ and the three RBCK points lightest in pion mass with $L=2.4\,{\rm fm}$ \cite{RBCK}.
	}
	\label{tableRBCK}
	\renewcommand{\arraystretch}{1.2}
\begin{tabular}{ll||cc|cc} 
	 & & \multicolumn{2}{l|}{(a) statistical error} & \multicolumn{2}{l}{(b) systematic envelope}  \\
	\hline
	$g_A^0$ &                                             & 1.223 $\pm$ 0.007 & fitted & 1.189 $\dots$ 1.230 & fitted \\   
	$g_1$ &                                               & 2.83 $\pm$ 0.26 & fitted & 2.5 $\dots$ 6.5 & fitted \\
	$C^\text{SSE}(1 \units{GeV})$ & $(\mathrm{GeV^{-2}})$ & $-1.67$ $\pm$ 0.37 & elim. & $-4.7$ $\dots$ $-1.3$ & elim. \\     
	$f_\pi$ & $(\mathrm{MeV})$                            & 92.4 & fixed  & 86.2 $\dots$ 92.4 & scanned \\    
	$c_A$ &                                               & 1.5 & fixed & 1.12 $\dots$ 1.5 & scanned \\         
	$\Delta$ & $(\mathrm{MeV})$                           & 271.1 & fixed & 271 $\dots$ 293 & scanned \\            
	\hline
	$\chi^2/\text{d.o.f.}$ &                              & 0.2 & & 0.19 $\dots$ 0.46 & \\         
	\hline
	$c_A^2 g_1$ &                                         & 6.4 $\pm$ 0.6 & & 5.7 $\dots$ 8.2 & \\  
	$C^\text{HB}(m_\pi^\text{phys})$ & $(\mathrm{GeV^{-2}})$ & $-0.43$ $\pm$ 0.11 & & $-0.6$ $\dots$ $0.5$ &  \\ 
	\end{tabular}
\end{table*}

\subsection{Expanding in powers of $m_\pi$} \label{gasseexp}

The effective field theory framework with explicit $\Delta\,(1232)$ degrees of freedom offers a way of studying convergence properties of $\pi N$ chiral perturbation theory. We perform such a test using the ${\cal O}(\epsilon^3)$ non-relativistic SSE expression  which is able to describe the quark mass dependence of $g_A$ over a large range of pion masses. 
We expand the expression (\ref{gasse}) for $m_\pi < \Delta$ in powers of $m_\pi$, around the chiral limit. The resulting series is of the form: 
\begin{align}
g_A =&\; g_A^0 \, \left\{ 1 + \left[ {{{\alpha_2}}} \ln
\frac{m_\pi}{\lambda} + {{{\tilde{\alpha}_2}}} \ln
\frac{2 \Delta}{\lambda} + {{\beta_2}} \right] \, m_\pi^2 + {{\alpha_3}} \, \frac{m_\pi^3}{\Delta} \right.
\nonumber\\ 
&\; \left. + \sum_{n\geq2}^{\phantom{i}}{{\gamma_n}} \frac{m_\pi^{2n}}{\Delta^{2n-2}}\ln \frac{m_\pi}{2 \Delta}  + \sum_{n\geq2}^{\phantom{i}}{{\beta_n}} \frac{m_\pi^{2n}}{\Delta^{2n-2}}\right\} + {\cal O}(\epsilon^4) \nonumber \\
=&\;g_A^0 \, \left[1+ A\,m_\pi^2 \ln{\frac{m_\pi}{\lambda}} + B(\lambda)\, m_\pi^2 + C\, m_\pi^3 + D\, m_\pi^4 \ln{\frac{m_\pi}{\lambda}}+ E(\lambda)\, m_\pi^4 + \dots \right]
\label{expansion}
\end{align}
All terms can be mapped onto the $\pi N$ Heavy Baryon expansion, according to the decoupling theorem \cite{AC,OS}. The series in $m_\pi^n/\Delta^m$ in the last equation corresponds to ``integrating out'' explicit delta degrees of freedom. The intermediate delta contributions are embedded in a string of couplings appearing in the Heavy Baryon Lagrangian. We note that in Eq.\,(\ref{expansion}) the terms starting from $m_\pi^3$ come entirely from diagrams 6 - 8 in Fig.\,\ref{gadiags}. For $m_\pi < \Delta$ a detailed numerical analysis of Eq.\,(\ref{expansion}) therefore gives an impression of the systematic errors introduced by integrating out the leading-order effect of the $\Delta\,(1232)$ in the HB$\chi$PT expansion. 

\begin{figure}[t]
  \begin{center}
    \includegraphics*[width=0.9\textwidth]{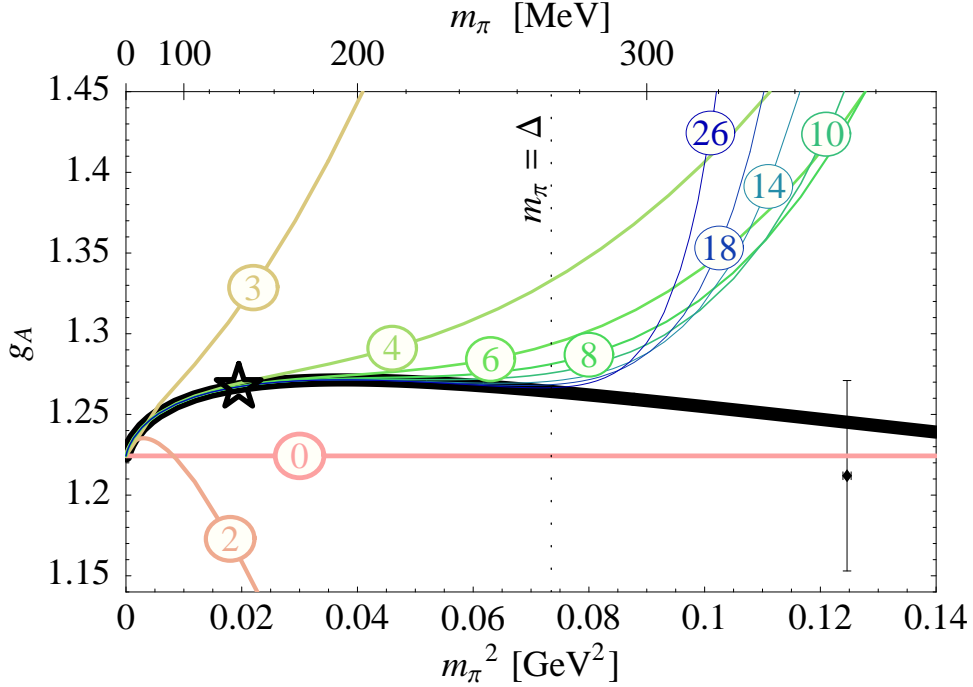}
    \caption{$g_A$ in SSE at order $\epsilon^3$: expansion in powers of $m_\pi$. Parameters have been set equal to the central values in the fit to the LHPC data. The labels $n$ of the curves give the order $m_\pi^n$ at which the series is truncated. The star denotes the physical point. The LHPC data point with smallest $m_\pi$ is also displayed.}
    \label{figconv}
  \end{center}
\end{figure}

\begin{figure}[h]
  \begin{center}
    \includegraphics*[width=0.8\textwidth]{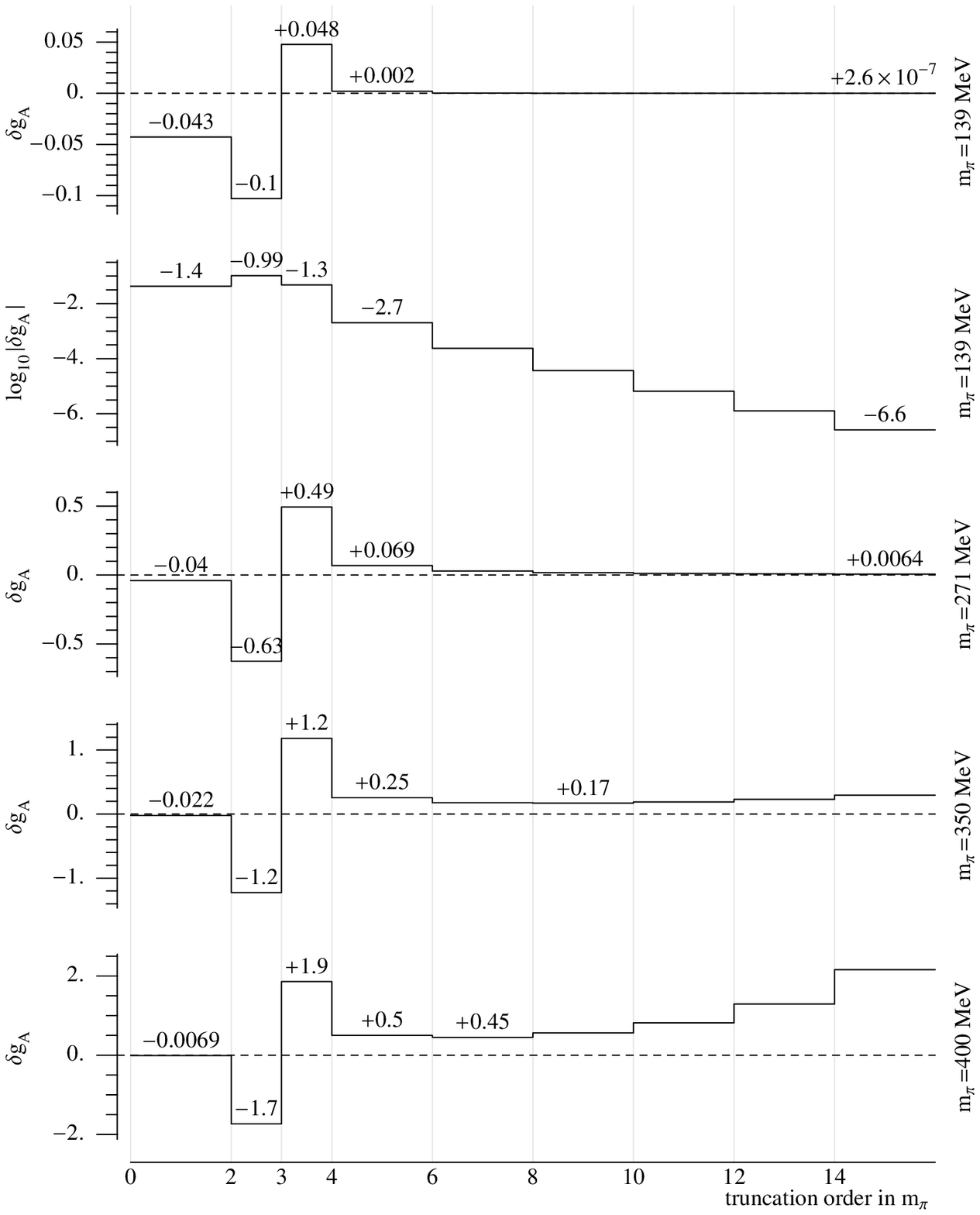}
    \caption{$g_A$ in SSE: convergence plot for the expansion of the ${\cal O}(\epsilon^3)$ result in powers of $m_\pi$. The dotted line refers to the full result with parameters set equal to the central values in the fit to the LHPC data.}
    \label{orders}
  \end{center}
\end{figure}

Fig.\,\ref{figconv} shows the convergence properties of the series (\ref{expansion}) towards the full ${\cal O}(\epsilon^3)$ result in Eq.\,(\ref{gasse}).
In Fig.\,\ref{orders} we plot the difference $\delta g_A$ between the expansion truncated at some power of $m_\pi$ and the full expression, for fixed values of the pion mass. In those plots we set $g_A^0$, $C^{\rm SSE}$ and $g_1$ equal to their central values in the fit to the LHPC data, see Table \ref{tableLHPC}.  

For small values of $m_\pi$ the expansion exhibits fast convergence. 
As soon as $m_\pi$ becomes larger than $\Delta$ we are outside the radius of convergence of the series in Eq.\,(\ref{expansion}). However, up to about $300\,{\rm MeV}$ in pion mass, the expansion truncated for example at $m_\pi^4$ still yields a good approximation. This is consistent with Ref.\,\cite{ulf} where a fifth-order counterterm proportional to $m_\pi^4$ turns out to be necessary and sufficient in the Heavy Baryon framework to get in contact with the smallest-$m_\pi$ lattice data. 

For $m_\pi > \Delta$ the expansion (\ref{expansion}) behaves like an asymptotic series \cite{asym}. For a given value of $m_\pi$ there is an optimal truncation order giving minimal deviation from the full result.  Beyond this order any truncation worsens the result. We find also that the optimal order shifts to lower powers when increasing the pion mass. At some value of $m_\pi$ the deviation, even at the best truncation order, exceeds the required level of precision.

Our analysis suggests that for $m_\pi \gtrsim 300\,{\rm MeV}$, once the pion mass exceeds the $\Delta N$ mass difference, the systematic errors in the Heavy Baryon expansion induced by removing the $\Delta\,(1232)$ as an explicit degree of freedom tend to get out of control. The series in Eq.\,(\ref{expansion}) is of course only part of the full Heavy Baryon result. It is thus in principle possible that the observed deviation $\delta g_A$ can be compensated by other higher order terms in the diagramatic expansion.

At this point it is worth noting that a reliable expression for the quark mass dependence of $g_A$ is crucial in order to consistently describe the quark mass dependence of the nucleon mass in the Heavy Baryon scheme at order $p^5$ \cite{birse1,birse2}. Our numerical analysis suggests that this cannot be achieved for $m_\pi \gtrsim 300\,{\rm MeV}$ in a chiral effective field theory scheme restricted to pion and nucleon degrees of freedom only.   

We have explored the sensitivity of our convergence study to $g_1$, which controls the relative strength of competing structures in the ${\cal O}(\epsilon^3)$ SSE formula (\ref{gasse}). We observed that for the rather ``large'' value $g_1 \approx 6$ an $m_\pi^4$-approximation is quite close to the SSE result for a large range of pion masses. We interpret this as a warning that an effective theory can spuriously appear to be well behaved at a specific truncation order for a particular artificial choice of parameters. This is in agreement with Ref.\,\cite{ulf} where the two-loop Heavy Baryon expression is found to describe the trend shown by lattice data up to $m_\pi \approx 600\,{\rm MeV}$ just for a particular choice of effective low-energy couplings.

Our analysis is based on the validity of the SSE perturbative framework. Our fits support the conclusion that in SSE important quark-mass dependent effects for $g_A$, involving the $\Delta(1232)$, are moved to low perturbative orders in the diagramatic expansion. However, before any firm conclusion can be drawn it is mandatory to study the effects of higher-order corrections and check the stability of our results. It is needless to say that for a numerical study at ${\cal O}(\epsilon^4)$, we must either wait for an improvement of the statistics for the $g_A$ lattice data or perform a combined analysis of $g_A$ and other nucleon observables characterized by a common subset of low-energy constants.


\section{Conclusions} \label{conclusions}
Summarizing the main results of this (continuum and infinite volume) analysis of $g_A$, we conclude as follows:

\begin{itemize}
\item Heavy Baryon Chiral Perturbation Theory restricted to pion and nucleon degrees of freedom only, when applied up to and including next-to-next-to-leading order, fails in attempts to produce meaningful interpolations between $g_A$ at the physical point and present lattice data. 
\item The explicit inclusion of the $\Delta\,(1232)$ at leading one-loop level, as implemented for example in the Small Scale Expansion scheme, is crucial in order to get a satisfactory description of the quark mass dependence of $g_A$ from the chiral limit across the physical point up to the lattice data. This does not come as a surprise in view of the well-known $\Delta$-dominance of $\pi N$ interactions and the Adler-Weisberger sum rule. 
\item An expansion of the leading one-loop expression with explicit $\Delta\,(1232)$ in powers of $m_\pi$ exhibits a stable  convergence pattern only for pion masses below $300\,{\rm MeV}$, the characteristic scale of the delta-nucleon mass difference. For larger values of $m_\pi$ the expansion behaves like an asymptotic series. Not unexpectedly, our analysis suggests that the systematic errors induced in the Heavy Baryon expansion by integrating out the $\Delta\,(1232)$, get out of control for $m_\pi \gtrsim 300\,{\rm MeV}$. 
\item  The addition of one heavier flavor in lattice simulations yields interpolation results that are statistically compatible with the two-flavor case.
\end{itemize}

\section{Acknowledgements}

We thank U.-G. Mei{\ss}ner, K. Orginos, D. B. Renner, D. Richards for helpful discussions. B.~M. acknowledges support by the DFG Emmy Noether-program. This work has been supported in part by BMBF and DFG.

\clearpage
\appendix
\vspace{2cm} \noindent {\huge\bf Appendix} 
\section{$d$-dimensional one-loop integrals} \label{integrals}

We encounter pion loop integrals in $d=4$ dimensions of the following basic type: 
\begin{align}
\Delta_\pi(m_\pi^2) =&\, \frac{1}{i}\int \frac{d^d k}{(2 \pi)^d} \frac{1}{m_\pi^2- k^2-i\epsilon}=m_\pi^{d-2}\, (4 \pi)^{-d/2}\,\Gamma\!\left(1-\frac{d}{2}\right)~.
\end{align}
Any ultraviolet divergence as $d$ tends to $4$ is subsumed in 
\begin{align} 
L(\lambda)=&\,\frac{\lambda^{d-4}}{16 \pi^2}\left\{\frac{1}{d-4}-\frac{1}{2}\left[\phantom{\frac{1}{1}}\!\!\!\!\! \ln{(4 \pi)}+\Gamma'(1)+1\right]\right\} \label{Ldiv}~,
\end{align}
where $\lambda$ is the dimensional regularization scale. Consequently,
\begin{align} 
\Delta_\pi(m_\pi^2) =&\,2 m_\pi^2\left[L(\lambda) + \frac{1}{16 \pi^2}\ln{\frac{m_\pi}{\lambda}}\right]~. 
\end{align}
\\
For loop integrals involving a nucleon propagator, consider the infrared singular part, cf. \cite{BL}, denoted by a subscript $I$ attached to the integral:
\begin{align}
I_N(p^2, m_\pi^2)&=\,\frac{1}{i}\int_I \frac{d^d k}{(2 \pi)^d}\frac{1}{(m_\pi^2-k^2 -i \epsilon)[M_0^2-(p-k)^2- i \epsilon]} \\ \nonumber \\
&=-\frac{p^2-M_0^2+m_\pi^2}{p^2}\, L(\lambda) +\bar{I}_N(p^2, m_\pi^2)~,\\ \nonumber \\
\bar{I}_N(p^2, m_\pi^2)&=-\frac{1}{8 \pi^2}\frac{\alpha \sqrt{1-\Omega^2}}{1+2 \alpha \Omega+ \alpha^2} \arccos{\left(-\frac{\Omega + \alpha}{\sqrt{1+2 \alpha \Omega + \alpha^2}}\right)}\nonumber \\ &\;\;\;\;-\frac{1}{16 \pi^2}\frac{\alpha(\alpha + \Omega)}{1+2 \alpha \Omega + \alpha^2}\left(2\ln{\frac{m_\pi}{\lambda}}-1\right)  \\ \nonumber \\
{\rm with\;\;\;\;\;}
\alpha&=\,\frac{m_\pi}{M_0}\,,\,\;\;\;\;\;\;\;\;\Omega=\frac{p^2-m_\pi^2-M_0^2}{2 m_\pi M_0}~.  \\ \nonumber
\end{align}
Furthermore,
\begin{align}
p^\mu I_N^{(1)}(p^2, m_\pi^2)=&\,\frac{1}{i}\int_I \frac{d^d k}{(2 \pi)^d}\frac{k^\mu}{(m_\pi^2-k^2 -i \epsilon)[M_0^2-(p-k)^2- i \epsilon]}~.\\ \nonumber
\end{align}
Using
\begin{equation}
p \cdot k = \frac{1}{2}(p^2 -M_0^2 +m_\pi^2)+\frac{1}{2}(k^2-m_\pi^2)-\frac{1}{2}\left[(p-k)^2-M_0^2\right] ~,
\end{equation}
one finds
\begin{align}
I_N^{(1)}(p^2, m_\pi^2)=&\,\frac{1}{2 p^2}\left[ (p^2-M_0^2+m_\pi^2)\, I_N(p^2, m_\pi^2)+ \Delta_\pi(m_\pi^2) \right]~.
\end{align}
\\
We use the following notations:
\begin{align}
I_N &\equiv I_N(p^2=M_0^2, m_\pi^2),\;\;\;\;\;\;\;\;\;\;\;I_N^{(1)} \equiv I_N^{(1)}(p^2=M_0^2, m_\pi^2),\\ \nonumber \\
I_\Delta(p^2) &\equiv I_N(p^2, M_0 \to M_\Delta^0),\;\;\;\;\;\;\;\;\;\;\;I_\Delta \equiv I_\Delta(p^2=M_0^2, m_\pi^2)~.
\end{align}

\newpage





\section{Details on the ${\cal O}(p^3)$ and ${\cal O}(p^4)$ calculations} \label{appgaBChPT}

At order $p^3$, nucleon field renormalization contributes in the way shown in Fig.\,\ref{gaZcube}.  
At this level of accuracy, the nucleon self-energy $\Sigma(p)$ is approximated as
\begin{equation}
\Sigma(p\!\!\!/=M_N) \approx \Sigma(p\!\!\!/=M_0)~,
\end{equation}
where $M_0$ is the nucleon mass in the $SU(2)$ chiral limit. This implies for the nucleon wave-function renormalization factor:
\begin{equation}
Z_N \approx 1+ \left.\frac{\partial \Sigma^{(3)}}{\partial p\!\!\!/}\right|_{p\!\!\!/=M_0}~.
\end{equation} 
Here $\Sigma^{(3)}$ is the nucleon self-energy at order $p^3$. Using infrared regularization \cite{BL},
\begin{equation}
\Sigma^{(3)}=\frac{3 g_A^0}{4 {f_\pi}^2} (M_0 +p\!\!\!/)\left[m_\pi^2\, I_N(p^2, m_\pi^2) + (M_0-p\!\!\!/)\,p\!\!\!/ \,I_N^{(1)}(p^2, m_\pi^2)\right]
\end{equation}
in terms of the basic integrals in Appendix \ref{integrals}.
Hence one finds:
\begin{align}
Z_N&=1-\frac{1}{32 \pi^2 f_\pi^2 M_0^3 \sqrt{4-m_\pi^2/M_0^2}}\left\{3{g_A^0}^2 m_\pi^2\left[(2m_\pi^3-6M_0^2 m_\pi)\arccos{\left(-\frac{m_\pi}{2M_0}\right)}\right. \right. \nonumber \\ &\left. \left. +\, M_0\sqrt{4-\frac{m_\pi^2}{M_0^2}}\left(M_0^2(1+48 \pi^2 L(\lambda))-32 \pi^2 L(\lambda) m_\pi^2+(3M_0^2-2m_\pi^2)\ln{\frac{m_\pi}{\lambda}}\right)\right] \right\}\nonumber\\& -8 B_{20}\, m_\pi^2 + 32\,F_2\, m_\pi^4~, \label{Zp3} 
\end{align}
where $L(\lambda)$ is defined in Eq.\,(\ref{Ldiv}).
Here $B_{20}$ is a third-order coupling \cite{HPW}, while $F_2$ enters in a fifth-order counterterm required to absorb the divergence proportional to $m_\pi^4$.

Projecting out the contributions to $g_A$ from the leading-one-loop amplitudes in Fig.\,\ref{diaggap3}, the following expression results, given in terms of the integrals in Appendix \ref{integrals}: 
\begin{align}
g_A&=g_A^0\,Z_N+ 4\,B_9\,m_\pi^2-\frac{g_A^0}{f_\pi^2} \Delta_\pi - 2\,\frac{g_A^0}{f_\pi^2} m_\pi^2 I_N + \frac{{g_A^0}^3}{4(d-1) f_\pi^2 M_0} \left[\phantom{\left. \frac{1}{1}\right|_{a}}\!\!\!\!\!\!\!\!\!2 m_\pi^2 M_0 (d-3) I_N \right. \nonumber \\ & \left.+ \left(2(d-3) M_0^2 m_\pi^2 +m_\pi^4 \right) \left.\frac{\partial}{\partial M_0}I_N(p^2)\right|_{p^2=M_0^2}+ (d-3)M_0 \Delta_\pi \right]+32\,F_1\, m_\pi^4~. \label{gap3impl}
\end{align}
$B_9$ is the third-order coupling already introduced in Ref.\,\cite{HPW} and $F_1$ takes care of an ultraviolet divergence. 

Combining  Eq.\,(\ref{Zp3}) and Eq.\,(\ref{gap3impl}), the result for the pion mass dependence of $g_A$ at order $p^3$ with infrared regularization is given by
\begin{align}
g_A=&\frac{1}{16 \pi^2 f_\pi^2 M_0^3 \sqrt{4- m_\pi^2/M_0^2}} \left\{g_A^0 m_\pi^3 (8({g_A^0}^2+1)M_0^2-(3 {g_A^0}^2+2)m_\pi^2) \arccos{\left(-\frac{m_\pi}{2 M_0}\right)}\nonumber \right. \\& \left.-M_0\sqrt{4-\frac{m_\pi^2}{M_0^2}}\left[M_0^2 m_\pi^2 {g_A^0}^3+(m_\pi^4 -16 f_\pi^2 M_0^2 \pi^2)g_A^0\nonumber \right. \right.\\& \left. \left. + ((4{g_A^0}^2+2)g_A^0 m_\pi^2 M_0^2 -(3{g_A^0}^2+2)g_A^0 m_\pi^4) \ln{\frac{m_\pi}{\lambda}}-64 C^r(\lambda) f_\pi^2 M_0^2 m_\pi^2 \pi^2\right]\right\}\nonumber \\ & + 32 F^r(\lambda) m_\pi^4~.  \label{gap3}
\end{align}
For the couplings involved in third order counterterms we use the notation already employed in our previous paper \cite{HPW}:
\begin{equation}
C^r(\lambda) \equiv B_9^r(\lambda)-2g_A^0 B_{20}^r(\lambda)~. \label{ccc}
\end{equation}
Furthermore, $F^r(\lambda)=F_1^r(\lambda)+F_2^r(\lambda)$. Those effective couplings are renormalized and regularization-scale-$\lambda$-dependent. They encode short-distance dynamics effects and scale in just such a way that the right-hand side of Eq.\,(\ref{gap3}) is scale independent:
\begin{align}
C^r(\lambda)&=B_9-2g_A^0 B_{20}-\frac{L(\lambda)}{f_\pi^2}\left(\frac{1}{2}g_A^0+{g_A^0}^3\right)\\
F^r(\lambda)&=F+\frac{L(\lambda)}{32 f_\pi^2 M_0^2}\,g_A^0(2+3{g_A^0}^2)~.
\end{align}

The factor 32 in the fifth-order counterterm in Eq.(\ref{gap3}) emphasizes that $F$ is the effective coupling to which one should apply ``naive" dimensional arguments, cf. Ref.\,\cite{GM}. 
Indeed the effective $\pi N$ Lagrangian at order $p^5$ can contribute via
\begin{equation}
{\cal{L}}_{\pi N}^{(5)}= 32\, f m_\pi^4\,\bar{\Psi} \,a_\mu \gamma^\mu \gamma_5 \Psi + \dots
\end{equation}
where we expect $f={\cal O}(1/\Lambda_\chi^4)$.

Expanding Eq.(\ref{gap3}) around $m_\pi=0$ leads to
\begin{align}
g_A=\;&g_A^0+\left[4\,C^r(\lambda)-\frac{{g_A^0}^3}{16 \pi^2 f_\pi^2}-\frac{g_A^0+2{g_A^0}^3}{8 \pi^2 f_\pi^2}\ln{\frac{m_\pi}{\lambda}}\right] m_\pi^2+ \frac{g_A^0+{g_A^0}^3}{8\pi f_\pi^2 M_0}m_\pi^3 \nonumber\\& +\left[32 F^r(\lambda)+\frac{g_A^0+2{g_A^0}^3}{16\pi^2 f_\pi^2 M_0^2}+\frac{g_A^0(2+3{g_A^0}^2)}{16 \pi^2 f_\pi^2 M_0^2}\ln{\frac{m_\pi}{\lambda}}\right] m_\pi^4+ {\cal{O}}(m_\pi^5)\,. \label{gap3chir} 
\end{align}
The sum of the first two terms in this expansion coincides with the leading-one-loop expression for $g_A$ in HB$\chi$PT, as expected in infrared regularization. The calculation gives a full tower of ``recoil corrections'' in the form of increasing powers of $1/M_0$. 
We include the counterterm $32\,F\,m_\pi^4$ in the third-order calculation in order to achieve renormalization without neglecting recoil corrections. Since contact terms {\it{up to and including}} ${\cal{O}}(p^3)$ cannot absorb higher-order divergences at $m_\pi^4$, their $\beta$-functions cannot compensate for scale dependence which is suppressed by two powers of $1/M_0$. By introducing the term $32\,F^r(\lambda)\,m_\pi^4$, we remove this unphysical scale dependence.\\

\begin{figure}[t]
  \begin{center}
    \includegraphics*[width=0.19\textwidth]{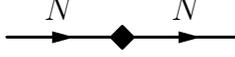}
     \caption{$c_1$-insertion in a nucleon line: the diamond corresponds to the vertex $i\,4\,c_1 m_\pi^2$ from ${\cal{L}}_{\pi N}^{(2)}$.}\label{c1}
  \end{center}
\end{figure}

At order $p^4$ nucleon field renormalization (Fig.\,\ref{gaZfourth}) and mass renormalization have been evaluated consistently within the accuracy in the perturbative, diagrammatic expansion at which we are working. In Fig.\,\ref{c1} we draw as a diamond the second order insertion in the nucleon line proportional to the low-energy constant $c_1$, see Eq.\,(\ref{Ldue}). 
At order $p^4$ we have to compute all graphs resulting from the insertion of {\it{at most one}} $c_1$-vertex in {\it{at most one}} nucleon line in the diagrams (2), (3) and (4) of Fig.\,\ref{diaggap3} and in the diagram of Fig.\,\ref{gaZcube}. Since for small $m_\pi$
\begin{equation}
\frac{i}{p\!\!\!/-(M_0-4\,c_1 m_\pi^2)}=\frac{i}{p\!\!\!/-M_0}+\frac{i}{p\!\!\!/-M_0}\,(i\,4\,c_1 m_\pi^2)\,\frac{i}{p\!\!\!/-M_0}+ \dots~,
\end{equation} 
we can summarize the $c_1$-insertions by a simple shift of the pole of the nucleon propagator from the ``bare'' nucleon mass to its renormalized value at second chiral order:
\begin{equation}
M_N=M_0-4\,c_1 m_\pi^2+{\cal{O}}(p^3)~.
\end{equation}
Therefore nucleon field renormalization is taken into account by
\begin{equation}
g_A^0 \left[1+\left.\frac{\partial}{\partial p\!\!\!/}\Sigma^{(3)}\right|_{p\!\!\!/ = M_0-4c_1 m_\pi^2} + \left.\frac{\partial}{\partial p\!\!\!/}\Sigma^{(4)}\right|_{p\!\!\!/ = M_0}\right]~. \label{uiu}
\end{equation}
Here $\Sigma^{(4)}(p\!\!\!/)$ is the contribution to the nucleon self-energy from the fourth-order tadpole in Fig.\,\ref{gaZfourth}, cf. \cite{BL},
\begin{equation}
\Sigma^{(4)} = \frac{3 m_\pi^2 \Delta_\pi}{f_\pi^0} \left( 2 c_1 - \frac{p^2}{M_0^2\,d} c_2 - c_3 \right)~.
\end{equation}
What is relevant at order $p^4$ is just the linear term in the expansion of Eq.\,(\ref{uiu}) in powers of $c_1$.

Each of the fourth-order diagrams in Fig.\,\ref{garelpfourth} contributes to $g_A$ as follows:
\begin{equation}
-\,\frac{g_A^0}{(d-1)f_\pi^2 M_0}\, m_\pi^2\left[\phantom{\frac{1}{1}}\!\!\!\! c_3 \left(\phantom{\frac{1}{1}}\!\!\!\!\!\Delta_\pi+(m_\pi^2-4M_0^2)\,I_N\right)+\,c_4\left(\phantom{\frac{1}{1}}\!\!\!\!\!\Delta_\pi+4 (d-2)M_0^2\,I_N+ m_\pi^2\,I_N\right)\phantom{\frac{1}{1}}\!\!\!\!\! \right]~. 
\end{equation}

The pion mass dependence of $g_A$ at order $p^4$ is finally given by
\newcommand{\gA}[1]{{g_A^0}^{#1}}
\newcommand{\fpi}{f_\pi}
\newcommand{\mpi}{m_\pi}
\newcommand{\MZ}{M_0}
\begin{align}
g_A 
= &\, \gA{} + 4\, C^r(\lambda)\, \mpi^2 
+ 32\,F^r(\lambda)\,m_\pi^4 + 128\,G^r(\lambda)\,m_\pi^6
- \, \frac{\gA{3}\mpi^2}{16 \pi^2 \fpi^2} 
\bigg[
  1-\frac{32 c_1 \MZ^2 \mpi^4 - 12 c_1 \mpi^6}{(4 \MZ^2 - \mpi^2)\, \MZ^3}
  \bigg] \nonumber\\ 
+ &\, \ln \left( \frac{\mpi}{\lambda} \right) \, \frac{\gA{}\mpi^2}{16 \pi^2 \fpi^2} 
\bigg[
  -2(1+2\gA{2})
  -(3 c_2 + 4 c_3 - 4 c_4) \frac{\mpi^2}{\MZ}
  +(2 + 3\gA{2}) \frac{\mpi^2}{\MZ^2} \nonumber \\ &
  +\frac{2}{3}(24 c_1 + c_3 + c_4 + 36 c_1 \gA{2}) \frac{\mpi^4}{\MZ^3}
  \bigg] \nonumber\\ 
+ &\, \arccos \left( -\,\frac{\mpi}{2 M_0} \right) \, \frac{\gA{} \mpi^3}{16 \pi^2 \fpi^2\,(4 \MZ^2 - \mpi^2)^{3/2}} 
\bigg[
  -\frac{128}{3}(c_3 - 2 c_4) \MZ^3 \nonumber \\ &
  +32 (1+\gA{2}) (\MZ^2 + 4 c_1 \MZ \mpi^2) 
  +32 (c_3 - c_4) \MZ \mpi^2
  -4(4+5\gA{2}) \mpi^2 \nonumber \\ &
  -8(12 c_1 + c_3 + 18 c_1 \gA{2}) \frac{\mpi^4}{\MZ}
  +(2+3\gA{2}) \frac{\mpi^4}{\MZ^2} 
  +2(24 c_1 + c_3 + c_4 + 36 c_1 \gA{2}) \frac{\mpi^6}{3 \MZ^3}
  \bigg] \nonumber\\ 
+ &\, \frac{\gA{} \mpi^4}{16 \pi^2 \fpi^2\,\MZ^3} 
\bigg[
	\frac{1}{12} (9 c_2 + 32 c_3 - 16 c_4) \MZ^2 -\MZ -\frac{5}{9} (c_3 + c_4) \mpi^2 
	\bigg] 
\label{gap4}
\end{align}
$G^r(\lambda)$ is a seventh-order effective coupling appearing in the counterterm required to absorb the divergence at $m_\pi^6$. This compensates the unphysical scale dependence at this power in $m_\pi$. The factor 128 is motivated by the fact that the effective $\pi N$ Lagrangian at order $p^7$ can contribute via
\begin{equation}
{\cal{L}}_N^{(7)}= 128\,g\, m_\pi^6 \,\bar{\Psi}\,\frac{\tau^i}{2} a_\mu^{i}\gamma^\mu \gamma_5 \Psi + \dots
\end{equation}
``Naive" dimensional arguments suggest $g ={\cal O}\left( 1/ \Lambda_\chi^6\right)$.
At order $p^4$ one finds
\begin{align}
F^r(\lambda)&=F+\frac{L(\lambda)}{32 f_\pi^2 M_0^2}\,(3{g_A^0}^3+2g_A^0-3c_2 g_A^0 M_0-4c_3 g_A^0 M_0+4c_4 g_A^0 M_0)\\
G^r(\lambda)&=G+\frac{L(\lambda)}{128 f_\pi^2 M_0^3}\,(24 c_1 {g_A^0}^3+16 c_1 g_A^0 +\frac{2}{3} c_3 g_A^0+\frac{2}{3} c_4 g_A^0)~.
\end{align}


\end{document}